\documentstyle[12pt]{article}

\renewcommand{\theequation}{\thesection\arabic{equation}}

\author{\sc
       P.M. Lavrov\thanks{E-mail: lavrov@tspu.edu.ru},\,\,P.Yu. Moshin
       \\
       \normalsize\it
       Tomsk State Pedagogical University, Tomsk 634041, Russia}
\title{\LARGE\bf
       Superfield Lagrangian Quantization with Extended BRST Symmetry}
\date{}
\begin{document}
\maketitle

\begin{quotation}
\small
\noindent
 We consider possible superfield representations of extended
 BRST symmetry for general gauge theories within the principle
 of gauge-fixing based on a generating equation for the gauge
 functional. We examine admissible superfield choices for an
 extended antibracket and delta-operator with given algebraic
 properties and show that only one of these choices is compatible
 with the requirement of extended BRST symmetry realized in terms
 of supertranslations along Grassmann coordinates. We demonstrate
 that this realization leads to the gauge-independence of the
 $S$-matrix.
\end{quotation}

\section{Introduction}
 In the past few years two tendencies have made themselves
 manifest in the development of covariant quantization
 methods based on (extended) BRST symmetry \cite{BRST,antiBRST}:
 one is introduction of gauge with the help of equations imposed
 on gauge-fixing functionals \cite{BBD,3pl,mod3pl};
 the other is formulation of superfield quantization rules
 \cite{LMR,BBD2,L,GM}. Recently both these concepts have been combined
 within the modified superfield formalism \cite{ModSF}, which
 generalizes the previous studies \cite{BBD,LMR} dealing with
 different modifications of the BV quantization \cite{BV} for general
 gauge theories in the framework of the standard BRST symmetry
 \cite{BRST}.

 In \cite{BBD}, it was shown that the BV approach can be formulated
 in such a way that the entire gauge-fixing part of the quantum action
 is subject to a special generating equation analogous to the usual
 generating equation that determines the quantum action. It was
 demonstrated that this concept of gauge-fixing guarantees the
 independence of the vacuum functional under infinitesimal
 gauge variations, which, by virtue of the equivalence theorem
 \cite{equiv}, implies the gauge-independence of the $S$-matrix.

 In \cite{LMR}, a superfield form of the BV quantization rules was
 proposed. The variables of the BV approach were combined into superfields
 $\Phi^A(\theta)$ and superantifields $\Phi^*_A(\theta)$ with opposite
 Grassmann parities, defined in a superspace with one anticommuting
 coordinate $\theta$. A superfield representation of the usual antibracket
 $(\,\,,\,)$ and the operator $\Delta$ was found. The vacuum functional
 is given as a functional integral over $\Phi^A(\theta)$ and $\Phi^*_A(\theta)$.
 The components of superfields and superantifields are interpreted as the
 usual field-antifield variables $\phi^A$, $\phi^*_A$ as well as the
 Langrange multipliers $\lambda^A$ and the sources $J_A$ to the fields. The
 fact that the components of the supervariables play different roles in the
 BV approach leads to a difference in the treatment of $\Phi^A(\theta)$
 and $\Phi^*_A(\theta)$. Thus, to obtain the correspondence with
 the BV vacuum functional one has to impose the constraint $J_A=0$ which
 is introduced in a manifestly superfield form by adding an integration
 density depending on $\Phi^*_A(\theta)$. In the space of the supervariables,
 the transformations of BRST symmetry are realized in terms of variations
 induced by supertranslations along the Grassmann coordinate. The
 generators of supertranslations naturally possess the properties of
 nilpotency and anticommutativity with each other as well as with the
 delta-operator, and their properties of differentiating the antibracket
 are identical with the well-known property of $\Delta$. The superfield
 BRST transformations were shown to encode the gauge-independence of the
 vacuum functional and, consequently, that of the corresponding $S$-matrix.

 Finally, in \cite{ModSF} the superfield formalism \cite{LMR} was
 modified in such a way that the gauge-fixing quantum action $X$
 in the vacuum functional was required to satisfy a special generating
 equation,
\begin{eqnarray}
\label{ModSupX}
 \frac{1}{2}(X,X)-UX=i\hbar\Delta X,
\end{eqnarray}
 expressed in terms of the generator $U$ of supertranslations in the
 space of superfields $\Phi^A(\theta)$, and formally analogous to the
 equation \cite{LMR} for the quantum action $W$
\begin{eqnarray}
\label{ModSupS}
\frac{1}{2}(W,W)+VW=i\hbar\Delta W,
\end{eqnarray}
 containing the generator $V$ of supertranslations in the space
 of superantifields $\Phi^*_A(\theta)$. The above generating
 equations (\ref{ModSupX}) and (\ref{ModSupS}) make it possible
 to encode the gauge-independence of the $S$-matrix in terms of
 the transformations of BRST symmetry, realized as a combination
 of supertranslations along the Grassmann coordinate and anticanonical
 transformations generated by the antibracket,
\begin{eqnarray}
\label{ModSupTr}
 \delta\Phi^A(\theta)&=&\mu U\Phi^A(\theta)+(\Phi^A(\theta),X-W)\mu,\nonumber\\
 \delta\Phi^*_A(\theta)&=&\mu V\Phi^*_A(\theta)+(\Phi^*_A(\theta),X-W)\mu,
\end{eqnarray}
 where $\mu$ is a constant anticommuting parameter. The modified superfield
 formalism \cite{ModSF} is equivalent to the gauge-fixing procedure
 \cite{BBD} and contains the original superfield approach \cite{LMR} as
 a special case of solutions to the generating equation for the gauge
 functional $X$.

 Note that there exists an alternative approach \cite{BBD2} to superfield
 quantization of gauge theories, following from a superfield Hamiltonian
 formalism \cite{BBD2} for constrained dynamical systems in the framework
 of BRST symmetry.

 Consider the most important differences between the methods \cite{BBD2} and
 \cite{LMR,ModSF}. First, the set of supervariables required for the
 construction of the vacuum functional in the covariant formulation
 \cite{BBD2} is composed by field-antifield pairs\footnote{We assume the
 separation of variables in terms of Darboux coordinates. For illustrative
 purposes, we have replaced the original notations \cite{BBD2} by those
 of \cite{LMR}, also in terms of the usual field-antifield variables
 $\phi^A$, $\phi^*_A$ and their superpartners $\lambda^A$, $J_A$, despite
 these latter do not admit of the interpretation \cite{LMR} from the
 viewpoint of the BV method. Note that the variables $\Phi^*_A(\theta)$
 considered in \cite{BBD2} are {\it formally} related to those introduced
 in \cite{LMR} by transposition of the components $\phi^*_A$ and $J_A$.}
 $\Phi^A(\theta)$, $\Phi^*_A(\theta)$ with the {\it same} Grassmann parity.
 Second, the method \cite{BBD2} is formally a {\it direct} extension of the
 BV approach, with the field-antifield variables replaced by their
 superfield counterparts. As a consequence, at the component level, the
 corresponding antibracket and delta-operator \cite{BBD2} are extensions of
 the original BV objects, constructed with the help of the superpartners
 $\lambda^A$, $J_A$ of the usual field-antifield variables $\phi^A$,
 $\phi^*_A$. As another consequence, the vacuum functional is given by a
 functional integral over the superfields $\Phi^A(\theta)$ with the
 variables $\Phi^*_A(\theta)$ placed on a specified gauge hypersurface.

 Although essentially different at the superfield level, the proposals
 \cite{LMR,BBD2,ModSF} are equivalent to each other in the sense that they
 parameterize the BV vacuum functional after imposing appropriate
 restrictions on the solutions of the generating equations.

 As in the case of the standard BRST symmetry realized in the reviewed papers
 \cite{BBD,LMR}, there have been studies \cite{3pl,mod3pl,L} devoted to the
 implementation of these concepts of gauge-fixing \cite{BBD} and superfield
 quantization \cite{LMR} on the basis of extended BRST symmetry \cite{antiBRST}.

 In the framework of the triplectic \cite{3pl} and modified triplectic
 \cite{mod3pl} methods, the idea of imposing a generating equation on the
 gauge-fixing part of the quantum action was incorporated into the $Sp(2)$
 covariant formulation \cite{Sp(2)} of extended BRST symmetry for general
 gauge theories. One of the ingredients of the triplectic (also called
 ``completely anticanonical'') approach is to consider part of the
 antifields and Lagrange multipliers of the $Sp(2)$ covariant formalism
 \cite{Sp(2)} as anticanonically conjugated variables, which involves the
 corresponding redefinition of the original \cite{Sp(2)} extended antibracket
 and delta-operator, retaining, however, their essential algebraic properties.

 In \cite{L}, a superfield form of the $Sp(2)$ covariant quantization rules
 was proposed along the lines of the papers \cite{LMR}. The variables of
 the $Sp(2)$ covariant approach were combined into a set of superfields
 $\Phi^A(\theta)$ and supersources $\bar{\Phi}_A(\theta)$, with the same
 Grassmann parity, defined in a superspace with two anticommuting
 coordinates $\theta^a$. The components of the superfields and supersources
 are interpreted as the fields $\phi^A$, the antifields $\phi^*_{Aa}$,
 $\bar{\phi}_A$, the Lagrange multipliers $\pi^{Aa}$, $\lambda^A$, and the
 sources $J_A$. The transformations of extended BRST symmetry are realized
 in terms of supertranslations in superspace with the corresponding
 generators $U^a$ (in the space of superfields) and $V^a$ (in the space of
 supersources) possessing the properties of generalized nilpotency and
 anticommutativity. The algebraic properties of the operators $U^a$ and
 $V^a$ with respect to the extended antibracket $(\,\,,\,)^a$ and the
 operator $\Delta^a$ naturally generalize the corresponding properties of
 the operators $U$ and $V$ with respect to the objects $(\,\,,\,)$ and
 $\Delta$ of the superfield quantization rules \cite{LMR,ModSF}.

 As compared to the superfield procedures \cite{LMR,ModSF},
 where the component representation of the antibracket and
 delta-operator is identical with the corresponding objects of the
 BV formalism, the superfield procedure \cite{L} applies an extended
 antibracket $(\,\,,\,)^a$ and operator $\Delta^a$ whose component
 representation turns out to coincide with the objects of the completely
 anticanonical approach \cite{3pl,mod3pl}, rather than with their original
 counterparts of the $Sp(2)$ covariant scheme \cite{Sp(2)}.

 Note that within the superfield methods \cite{LMR,ModSF}, based on the
 standard BRST symmetry, where the superspace contains a single
 anticommuting coordinate, there exists only one possibility of
 constructing the antibracket $(\,\,,\,)$ and the operator $\Delta$ with
 the algebraic properties \cite{LMR,ModSF}, under such natural requirements
 as a specific Grassmann parity and locality in $\theta$. In this sense,
 there is a unique realization of the superfield rules \cite{LMR,ModSF},
 given the set of supervariables, the general form of the vacuum functional
 and the generating equations.

 In the case of a superfield formalism based on the extended BRST symmetry,
 the situation appears to be more complicated, because a superspace with
 two Grassmann coordinates admits different possibilities of constructing
 objects with the properties \cite{Sp(2)} of the extended antibracket
 $(\,\,,\,)^a$ and the operator $\Delta^a$. Thus, in the recent paper
 \cite{GM}, a superfield form of the $osp(1,2)$ covariant quantization
 rules \cite{osp} was proposed, where another realization of superfield
 objects with the given properties was found. Moreover, the component form
 of the these objects is identical with the original extended antibracket
 $(\,\,,\,)^a$ and the operator $\Delta^a$, as defined in \cite{Sp(2)}.

 In this paper we investigate possible superfield representations of
 extended BRST symmetry for general gauge theories within the principle of
 gauge-fixing based on a generating equation for the gauge functional.

 To this end, we generalize the superfield $Sp(2)$ covariant scheme
 \cite{L} along the lines of the modified superfield approach \cite{ModSF},
 taking into account the arbitrariness in a specific choice of its basic
 objects. Thus, we postulate generating equations for the superfiled
 quantum action $S(\Phi,\bar{\Phi})$ and the gauge-fixing functional
 $X(\Phi,\bar{\Phi})$ in a form analogous to eqs.~(\ref{ModSupX}) and
 (\ref{ModSupS}), i.e. expressed in terms of an extended antibracket
 $(\,,\,)^a$ and operators $\Delta^a$, $U^a$, $V^a$. While the manifest
 form of the operators $U^a$ and $V^a$ is determined by their
 interpretation as generators of supertranslations, there is still an
 ambiguity in the choice of the extended antibracket $(\,\,,\,)^a$ and the
 operator $\Delta^a$ with the given algebraic properties \cite{L}.

 In this connection, we consider possible representations of these objects,
 and observe that they are reduced to two admissible choices, whose
 component form in one case coincides with the extended antibracket and
 delta-operator of the original $Sp(2)$ covariant scheme \cite{Sp(2)}, and
 in the other case, with their counterparts of the triplectic approach
 \cite{3pl,mod3pl}. Defining the vacuum functional as a straightforward, in
 the sense of \cite{ModSF}, extension of the vacuum functional \cite{L}, we
 demonstrate that only one of the two choices for the objects $(\,\,,\,)^a$
 and $\Delta^a$, namely, the one applied by the completely anticanonical
 procedure \cite{3pl,mod3pl}, is compatible with the requirement of
 extended BRST symmetry considered as a generalization of
 eqs.~(\ref{ModSupTr}) in the form of variations of superfields
 $\Phi^A(\theta)$ and supersources $\bar{\Phi}_A(\theta)$ induced by
 supertranslations combined with anticanonical transformations generated by
 the extended antibracket $(\,\,,\,)^a$. Furthermore, we show that the
 postulated form of extended BRST symmetry encodes the gauge-independence
 of the $S$-matrix.

 We demonstrate that, on the one hand, the resulting quantization rules
 provide a superfield form of the modified triplectic approach
 \cite{mod3pl}, and, on the other hand, they contain the superfield $Sp(2)$
 covariant scheme \cite{L} as a particular case of solutions for the
 gauge-fixing functional $X$.

 The paper is organized as follows. In Section 2 we introduce the basic
 objects $U^a$, $V^a$, $\Delta^a$, $(\,\,,\,)^a$ with the algebraic
 properties \cite{L} and give their manifest superfield representation. In
 Section 3 we extend the quantization rules \cite{L} along the lines of
 \cite{ModSF} and determine the choice of $\Delta^a$ and $(\,\,,\,)^a$
 compatible with the given form of extended BRST symmetry. In Section 4 we
 demonstrate the gauge-independence of the $S$-matrix in the proposed
 superfield formalism. In Section 5 we discuss the connection of our
 formalism with the methods \cite{mod3pl,L}. In Appendix we analyze the
 properties of the operators $U^a$, $V^a$ and $\Delta^a$ from the viewpoint
 of Lie superalgebras.

 We use the condensed notations \cite{DeW} and the conventions adopted in
 \cite{L}.

\section{Main Definitions}
\setcounter{equation}{0}

 Consider a superspace ($x^\mu$, $\theta^a$), where $x^\mu$ are space-time
 coordinates, and $\theta^a$ is an $Sp(2)$ doublet of anticommuting
 coordinates. Note that any function $f(\theta)$ has a component
 representation,
\[
 f(\theta)=f_0+\theta^a f_a+\theta^2 f_3,\,\,\,
 \theta^2\equiv\frac{1}{2}\theta_a\theta^a,
\]
 and an integral representation,
\[
 f(\theta)=\int d^2\,\theta'\,\delta(\theta'-\theta)f(\theta'),\,\,\,
 \delta(\theta'-\theta)=(\theta'-\theta)^2,
\]
 where raising and lowering the $Sp(2)$ indices is performed by the rule
 $\theta^a=\varepsilon^{ab}\theta_b$, $\theta_a=\varepsilon_{ab}\theta^b$,
 with $\varepsilon^{ab}$ being a constant antisymmetric tensor,
 $\varepsilon^{12}=1$, and integration over $\theta^a$ is given by
\[
\int d^2\theta=0,\;\;\int d^2\theta\;\theta^a=0,\;\;\int d^2\theta\;
\theta^a\theta^b=\varepsilon^{ab}.
\]
 In particular, for any function $f(\theta)$ we have
\[
 \int d^2\theta\;\frac{\partial f(\theta)}{\partial \theta^a}=0,
\]
 which implies the property of integration by parts
\[
\int d^2\theta\;\frac{\partial f(\theta)}{\partial \theta^a}g(\theta)=
-\int d^2 \theta (-1)^{\varepsilon(f)}f(\theta)\frac{\partial g(\theta)}
{\partial\theta^a}\,,
\]
 where derivatives with respect to $\theta^a$ are taken from the left.

 According to \cite{L}, we now introduce a set of superfields
 $\Phi^A{(\theta)}$, $\varepsilon(\Phi^A)=\varepsilon_A$, with the boundary
 condition
\[
 \left.\Phi^A(\theta)\right|_{\theta=0}=\phi^A
\]
 and a set of supersources $\bar{\Phi}_A{(\theta)}$ of the same Grassmann
 parity, $\varepsilon(\bar{\Phi}_A)=\varepsilon_A$.

 Denote by $U^a$ and $V^a$ doublets of Fermionic operators \cite{L}
 generating transformations of superfields and supersources induced
 by supertranslations $\theta^a\to\theta^a+\mu^a$ along the Grassmann
 coordinates,
\begin{eqnarray*}
\delta\Phi^A(\theta)&=&\mu_a\frac{\partial \Phi^A(\theta)}
{\partial \theta_a}=\mu_aU^a\Phi^A(\theta),\\
\delta\bar{\Phi}_A(\theta)&=&\mu_a\frac{\partial\bar{\Phi}_A(\theta)}
{\partial \theta_a}=\mu_aV^a\bar{\Phi}_A(\theta).
\end{eqnarray*}
 The generators $U^a$ and $V^a$ can be represented as first-order
 differential operators, having the form of $\theta$-local functionals,
\begin{eqnarray}
 \label{U&V}
 U^a&=&\int d^2\theta\frac {\partial\Phi^A(\theta)}{\partial\theta_a}
 \frac {\delta_l}{\delta\Phi^A(\theta)},\nonumber\\
 V^a&=&\int d^2\theta\frac {\partial\bar{\Phi}_A(\theta)}{\partial
 \theta_a}\frac {\delta}{\delta\bar{\Phi}_A(\theta)},
\end{eqnarray}
 where
\begin{eqnarray*}
\nonumber
 &&\frac{\delta_l\Phi^A(\theta)}{\delta\Phi^B(\theta^{'})}
 =\delta(\theta^{'}-\theta)\delta^A_B
 =\frac{\delta\Phi^A(\theta)}{\delta\Phi^B(\theta^{'})},\\
\nonumber
 &&\frac{\delta\bar{\Phi}_A(\theta)}{\delta\bar{\Phi}_B(\theta^{'})}
 =\delta(\theta^{'}-\theta)\delta^B_A.
\end{eqnarray*}
 From eqs.~(\ref{U&V}) follow the algebraic properties
\begin{eqnarray}
\label{algUV}
 U^{\{a}U^{b\}}=0,\;\;V^{\{a}V^{b\}}=0,\;\;V^aU^b+U^bV^a=0.
\end{eqnarray}

 Let us introduce a doublet of Fermionic second-order differential
 operators $\Delta^a$ with the properties of generalized nilpotency
 and anticommutativity \cite{L}
\begin{eqnarray}
\label{algDelta}
 &&\Delta^{\{a}\Delta^{b\}}=0,\nonumber\\
 &&\Delta^{\{a}V^{b\}}+V^{\{a}\Delta^{b\}}=0,\nonumber\\
 &&\Delta^{\{a}U^{b\}}+U^{\{a}\Delta^{b\}}=0,
\end{eqnarray}
 where the curly brackets denote symmetrization over $Sp(2)$
 indices, $A^{\{a}B^{b\}}=A^aB^b+A^bB^a$.

 Note that the relations (\ref{algUV}), (\ref{algDelta}) imposed
 on arbitrary linearly independent Fermionic doublets $U^a$, $V^a$,
 $\Delta^a$ define a set of nilpotent Lie superalgebras ${\cal G}$,
 with $6\leq {\rm dim}\,{\cal G}\leq 8$ (see Appendix).

 The action of $\Delta^a$ on the product of any two
 functionals $F$, $G$ {\it defines} an antibracket operation
 $(\,\,,\,)^a$
\begin{eqnarray}
\label{antibr}
 \Delta^a(F\cdot G)=(\Delta^aF)\cdot G+F\cdot(\Delta^a G)(-1)^{\varepsilon(F)}
 +(F,G)^a(-1)^{\varepsilon(F)}
\end{eqnarray}
 with the properties
\begin{eqnarray}
\label{antibprop}
 &&\varepsilon((F,G)^a)=\varepsilon(F)+\varepsilon(G)+1,\nonumber\\
 &&(F,G)^a=-(-1)^{(\varepsilon (F)+1)(\varepsilon (G)+1)}(G,F)^a,\nonumber\\
 &&D^{\{a}(F,G)^{b\}}=(D^{\{a}F,G)^{b\}}-(F,D^{\{a}G)^{b\}}
 (-1)^{\varepsilon(F)},\\
 \nonumber\\
\label{fgh}
 &&(F,GH)^a=(F,G)^aH+(F,H)^aG(-1)^{\varepsilon(G)\varepsilon(H)},\\
 \nonumber\\
\label{Jacobi}
 &&((F,G)^{\{a},H)^{b\}}(-1)^{(\varepsilon (F)+1)(\varepsilon (H)+1)}
 +{\rm cycle}\,(F,G,H)\equiv 0,
\end{eqnarray}
 where $D^a=(\Delta^a,U^a,V^a)$. Note that eqs.~(\ref{antibprop}) follow
 immediately from the definition (\ref{antibr}) and the relations
 (\ref{algDelta}). Eq.~(\ref{fgh}) is the consequence of the fact that
 $\Delta^a$ in eq.~(\ref{antibr}) is assumed to be a second-order
 differential operator, while the generalized Jacobi identity (\ref{Jacobi})
 follows from eqs.~(\ref{algDelta})--(\ref{fgh}).

 Finally, in terms of $U^a$, $V^a$ and $\Delta^a$ we define the operators
\begin{eqnarray}
\label{Deltas_nExSQ}
 \bar{\Delta}^a=\Delta^a + \frac{i}{\hbar}V^a,\;\;
 \tilde{\Delta}^a=\Delta^a - \frac{i}{\hbar}U^a
\end{eqnarray}
 with the properties
\[
 \bar{\Delta}^{\{a}\bar{\Delta}^{b\}}=0,\;
 \tilde{\Delta}^{\{a}\tilde{\Delta}^{b\}}=0,\;
 \bar{\Delta}^{\{a}\tilde{\Delta}^{b\}} +
 \tilde{\Delta}^{\{a}\bar{\Delta}^{b\}}=0
\]
 and
\begin{eqnarray*}
 \bar{\Delta}^{\{a}(F,G)^{b\}}&=&(\bar{\Delta}^{\{a}F,G)^{b\}}
 -(F,\bar{\Delta}^{\{a}G)^{b\}}(-1)^{\varepsilon(F)},\nonumber\\
 \tilde{\Delta}^{\{a}(F,G)^{b\}}&=&(\tilde{\Delta}^{\{a}F,G)^{b\}}
 -(F,\bar{\Delta}^{\{a}G)^{b\}}(-1)^{\varepsilon(F)},
\end{eqnarray*}
 following from eqs.~(\ref{algUV}), (\ref{algDelta}) and (\ref{antibprop}).

 An explicit form of the extended operator $\Delta^a$ and the corresponding
 extended antibracket $(\,\,,\,)^a$ with the given properties is not {\it
 unique}.

 Consider the class of Fermionic second-order differential operators (in
 the space of superfields and supersources) such that the dependence on the
 components of the $\Phi^A(\theta)$ and $\bar{\Phi}_A(\theta)$ enters only
 through the derivatives
\[
 \frac{\delta}{\delta\Phi^A(\theta)},\,\,
 \frac{\delta}{\delta\bar{\Phi}_A(\theta)}.
\]
 In the specified class there exist only two linearly independent $Sp(2)$
 doublets having the form of $\theta$-local functionals and possessing the
 algebraic properties (\ref{algDelta}) of the extended
 delta-operator.\footnote{In all, the specified class contains four linearly
 independent $Sp(2)$ doublets having the form of $\theta$-local operator
 functionals. The possibilities are limited to integrands constructed from
 the functional derivatives $\frac{\delta}{\delta\Phi^A(\theta)}$,
 $\frac{\delta}{\delta\bar{\Phi}_A(\theta)}$ and different combinations of
 $\theta^a$, $\frac{\partial}{\partial\theta^a}$. Note that the analysis of
 such combinations is simplified due to integration by parts and the use of
 the anticommutator
 $\{\theta^a,\frac{\partial}{\partial\theta^b}\}=\delta^a_b$.} Due to an
 additional property of anticommutativity with each other, these operators
 span a two-dimensional linear space of operators with the properties of
 $\Delta^a$. The basis elements $\Delta^a_1$, $\Delta^a_2$ of this space
 can be chosen in the form
\begin{eqnarray}
\label{DeltaaExSQ}
 &\!\!\!\!\!\!\!\!\!\!\!
 &\Delta^a_1=-\int d^2\theta \frac {\delta_{\it l}}{\delta\Phi^A(\theta)}
 \frac {\partial}{\partial \theta_a}\frac
 {\delta}{\delta\bar{\Phi}_A(\theta)}\,,\\
 &\!\!\!\!\!\!\!\!\!\!\!\!\!\!\!\!\!\!\!
\label{DeltaaExSQ2}
 &\Delta^a_2=\int d^2\theta \frac {\delta_{\it l}}{\delta\Phi^A(\theta)}
 \frac{\partial^2}{\partial\theta^2}\left(\theta^a
 \frac{\delta}{\delta\bar{\Phi}_A(\theta)}\right)\!,
\end{eqnarray}
 where
\[
 \frac{\partial^2}{\partial\theta^2}\equiv\frac{1}{2}\varepsilon^{ab}
 \frac{\partial}{\partial\theta^b}\frac{\partial}{\partial\theta^a}.
\]
 The extended delta-operators (\ref{DeltaaExSQ}) and (\ref{DeltaaExSQ2})
 generate the corresponding extended antibrackets
\begin{eqnarray}
\label{ABExSQ}
 &\!\!\!\!\!\!\!\!\!\!\!
 &(F,G)^a_1=\int d^2\theta\Bigg\{\frac{\delta F}{\delta\Phi^A(\theta)}
 \frac{\partial}{\partial\theta_a}\frac{\delta G}
 {\delta\bar{\Phi}_A(\theta)}(-1)^{\varepsilon_A+1}
 -(F\leftrightarrow G)(-1)^{(\varepsilon (F)+1)
 (\varepsilon (G)+1)}\Bigg\}
\end{eqnarray}
 and
\begin{eqnarray}
\label{ABExSQ2}
 &\!\!\!\!\!\!\!\!\!\!\!\!\!\!\!\!\!\!\!\!
 &(F,G)^a_2=\int d^2\theta\Bigg\{\!\!\left(\frac{\partial^2}{\partial\theta^2}
 \frac{\delta F}{\delta\Phi^A(\theta)}\right)\theta^a
 \frac{\delta G}{\delta\bar{\Phi}_A(\theta)}(-1)^{\varepsilon_A}
 -(F\leftrightarrow G)(-1)^{(\varepsilon (F)+1)
 (\varepsilon (G)+1)}\!\Bigg\}.
\end{eqnarray}

 The choice of the operator $\Delta^a$ and the antibracket $(\,\,,\,)^a$
 in the form (\ref{DeltaaExSQ}), (\ref{ABExSQ}) is identical with
 the one used in \cite{L}, whereas the other choice, in the form
 (\ref{DeltaaExSQ2}), (\ref{ABExSQ2}), was made in \cite{GM}.

 Despite the fact that the specific representations (\ref{DeltaaExSQ}),
 (\ref{DeltaaExSQ2}) of the operator $\Delta^a$ both satisfy
 eqs.~(\ref{algDelta}), they are nevertheless not equivalent at the
 algebraic level. Namely, it can be shown that these choices select two
 different Lie superalgebras associated with the whole set of relations
 (\ref{algUV}), (\ref{algDelta}) for the operators $U^a$, $V^a$, $\Delta^a$
 (see Appendix).

\section{Quantization Rules. Extended BRST Symmetry}
\setcounter{equation}{0}

 Consider a generalization of the superfield $Sp(2)$ covariant
 quantization rules \cite{L} modified along the lines of \cite{ModSF}.
 Define the vacuum functional $Z$ as the following path integral:
\begin{eqnarray}
\label{ZExSQ}
 Z=\int d\Phi\, d\bar{\Phi}\,\rho (\bar{\Phi})\exp \left\{
 \frac{i}{\hbar}\bigg[W(\Phi,\bar{\Phi})+ X(\Phi,\bar{\Phi})
 + \bar{\Phi}\Phi\bigg]\right\},
\end{eqnarray}
 where $W=W(\Phi,\bar{\Phi})$ is a quantum action that satisfies
 the generating equation
\begin{eqnarray}
\label{GEqExSQ}
 \bar{\Delta}^a\exp\bigg\{\frac{i}{\hbar}W\bigg\}=0,
\end{eqnarray}
 and $X=X(\Phi,\bar{\Phi})$ is a Bosonic gauge-fixing functional
 subject to the equation
\begin{eqnarray}
\label{GEqXExSQ}
 \tilde{\Delta}^a\exp\bigg\{\frac{i}{\hbar}X\bigg\}=0,
\end{eqnarray}
 with $\bar{\Delta}^a$ and $\tilde{\Delta}^a$ defined by
 eqs.~(\ref{Deltas_nExSQ}). Eqs.~(\ref{GEqExSQ}) and (\ref{GEqXExSQ})
 are equivalent to
\begin{eqnarray}
\label{GEq1ExSQ}
 \frac{1}{2}(W,W)^a+V^aW&=&i\hbar\Delta^aW,\\
\label{GEqX1ExSQ}
 \frac{1}{2}(X,X)^a - U^aX&=&i\hbar\Delta^aX.
\end{eqnarray}
 In eq.~(\ref{ZExSQ}), we have used the notation $\rho(\bar{\Phi})$
 for a functional which defines the weight of integration over the
 supersources $\bar{\Phi}_A(\theta)$ and has the form of a functional $\delta$
 function,
\begin{eqnarray}
\label{WegFExSQ}
 \rho(\bar{\Phi})=\delta\left(\int d^2\theta\,\bar{\Phi}(\theta)\right)\!.
\end{eqnarray}
 We have also introduced the functional
\begin{eqnarray}
\label{BilFExSQ}
 \bar{\Phi}\Phi\equiv\int d^2\theta\;\bar{\Phi}_A(\theta)\Phi^A(\theta).
\end{eqnarray}

 Define the transformations of extended BRST symmetry
 as the following transformations of global supersymmetry:
\begin{eqnarray}
\label{BRSTExSQ}
 && \delta\Phi^A(\theta)=\mu_a U^a\Phi^A(\theta) + (\Phi^A(\theta),
 X - W)^a\mu_a,
 \nonumber\\
 &&\delta\bar{\Phi}_A(\theta)=\mu_a V^a\bar{\Phi}_A(\theta) +
 (\bar{\Phi}_A(\theta), X - W)^a\mu_a.
\end{eqnarray}

 On the one hand, eqs.~(\ref{BRSTExSQ}) provide a straightforward
 generalization of eqs.~(\ref{ModSupTr}) introduced in \cite{ModSF}.
 One the other hand, the choice of extended BRST symmetry in
 this particular form is motivated by the fact that in case the
 transformations (\ref{BRSTExSQ}) do realize the invariance of the
 integrand (\ref{ZExSQ}), then in combination with the generating
 equations (\ref{GEqExSQ}) and (\ref{GEqXExSQ}) they also guarantee
 the gauge-independence of the vacuum functional.

 Note that prior to introducing the symmetry transformations
 (\ref{BRSTExSQ}) the explicit choice of the extended antibracket
 $(\,\,,\,)^a$ in the form (\ref{ABExSQ}) has no clear advantage over the
 other choice (\ref{ABExSQ2}). Let us show, however, that
 eq.~(\ref{ABExSQ}) does meet the above requirement of extended BRST
 symmetry, while eq.~(\ref{ABExSQ2}) does not.

 Consider the change of the integrand in eq.~(\ref{ZExSQ})
 under the transformations (\ref{BRSTExSQ}). To examine the
 change of the exponential in eq.~(\ref{ZExSQ}), note that
 the variation of an arbitrary functional $F$ under the
 transformations
\[
 \delta_{(1)}\Phi^A(\theta)=\mu_aU^a\Phi^A(\theta),\,\,\,
 \delta_{(1)}\bar{\Phi}_A(\theta)=\mu_aV^a\bar{\Phi}_A(\theta),
\]
 induced by supertranslations, has the form
\[
 \delta_{(1)}F=\mu_a(U^a+V^a)F.
\]
 In particular, for the functional $\bar{\Phi}\Phi$ (\ref{BilFExSQ})
 we have
\[
 \delta_{(1)}(\bar{\Phi}\Phi)=0.
\]
 At the same time, in both cases (\ref{ABExSQ}) and (\ref{ABExSQ2})
 of explicit representation of the extended antibracket the
 anticanonical transformations
\[
 \delta_{(2)}\Phi^A(\theta)=(\Phi^A(\theta),Y)^a\mu_a,\,\,\,
 \delta_{(2)}\bar{\Phi}_A(\theta)=(\bar{\Phi}_A(\theta),Y)^a\mu_a,\,\,\,
 \varepsilon(Y)=0
\]
 generate the corresponding transformations of an arbitrary functional
 $F$
\[
 \delta_{(2)}F=(F,Y)^a\mu_a.
\]
 As a result, eqs.~(\ref{BRSTExSQ}) lead to the variation $\delta=\delta_{(1)}+
 \delta_{(2)}$
\begin{eqnarray}
\label{generl}
 \delta(W+X+\bar{\Phi}\Phi)&=&\mu_a\bigg((W,W)^a-(X,X)^a+(U^a+V^a)(W+X)\bigg)+
                              \nonumber\\
                           &+&\mu_a(\bar{\Phi}\Phi,W-X)^a.
\end{eqnarray}
 To examine the change of the integration measure in eq.~(\ref{ZExSQ}), note
 that in both cases (\ref{ABExSQ}) and (\ref{ABExSQ2})
 the weight functional $\rho(\bar{\Phi})$ is invariant
 under the transformations (\ref{BRSTExSQ}), $\delta\rho(\bar{\Phi})=0$,
 while the corresponding Jacobian $J$ has the form
\begin{eqnarray}
\label{J}
 J = \exp (2\mu_a\Delta^a W - 2\mu_a\Delta^a X).
\end{eqnarray}

 Denote by $I$ the integrand in eq.~(\ref{ZExSQ}). Then, by
 virtue of eqs.~(\ref{GEq1ExSQ}), (\ref{GEqX1ExSQ}), (\ref{generl})
 and (\ref{J}), its variation $\delta I$ under the transformations
 (\ref{BRSTExSQ}) is given by
\begin{eqnarray}
\label{deltaI}
 \delta I=i\hbar^{-1}\mu_a I\bigg((U^a-V^a)(W-X)+(\bar{\Phi}\Phi,W-X)^a\bigg),
\end{eqnarray}
 and hence the condition of invariance of the integrand takes the form
\begin{eqnarray}
\label{equation}
 (U^a-V^a)(W-X)+(\bar{\Phi}\Phi,W-X)^a=0.
\end{eqnarray}

 The fulfillment of eq.~(\ref{equation}) obviously
 depends on a specific choice of the extended
 antibracket. Thus in the case of the antibracket
 (\ref{ABExSQ}) the above condition is satisfied due
 to the identity
\[
 (\bar{\Phi}\Phi,F)^a=(V^a-U^a)F,
\]
 which, according to eq.~(\ref{deltaI}), implies the invariance
 of the integrand in eq.~(\ref{ZExSQ}) under the transformations
 (\ref{BRSTExSQ}).

 On the other hand, in the case of the antibracket (\ref{ABExSQ2}) we have
\[
 (\bar{\Phi}\Phi,F)^a\not\equiv(V^a-U^a)F,
\]
 which means that eq.~(\ref{equation}) does not hold identically, and
 therefore the integrand is not invariant under the transformations
 (\ref{BRSTExSQ}) without additional restrictions on the functionals
 $W$ and $X$.

\section{Gauge Independence}
\setcounter{equation}{0}

 Let us study the dependence of the vacuum functional $Z$ (\ref{ZExSQ}) on
 the choice of gauge, using the explicit form of the extended antibracket
 (\ref{ABExSQ}) and the corresponding extended delta-operator
 (\ref{DeltaaExSQ}), which provide the invariance of the integrand
 (\ref{ZExSQ}) under the transformations (\ref{BRSTExSQ}).

 Note, first of all, that any admissible variation $\delta X$
 of the gauge functional $X$ must satisfy the equation
\[
 (X,\delta X)^a - U^a\delta X = i\hbar \Delta^a \delta X,
\]
 which can be represented in the form
\begin{eqnarray}
\label{VarXExSQ}
 \hat{Q}^a(X)\delta X=0.
\end{eqnarray}
 In eq.~(\ref{VarXExSQ}), we have introduced an operator $\hat{Q}^a(X)$
 possessing the property of generalized nilpotency,
\begin{eqnarray}
\label{QaExSQ}
\hat{Q}^a(X)=\hat{\cal
 B}^a(X)-i\hbar\tilde{\Delta}^a, \quad
\hat{Q}^{\{a}(X)\hat{Q}^{b\}}(X) = 0,
\end{eqnarray}
 where $\hat{\cal B}^a(X)$ stands for an operator
 acting by the rule
\[
 (X,F)^a\equiv\hat{\cal B}^a(X)F
\]
 and possessing the property
\[
 \hat{\cal B}^{\{a}(X)\hat{\cal B}^{b\}}(X)=
 \hat{\cal B}^{\{a}\left(\frac{1}{2}(X,X)^{b\}}\right)\!.
\]
 By virtue of the operator $\hat{Q}^a(X)$ in eqs.~(\ref{QaExSQ}),
 any functional
\begin{eqnarray}
\label{DeltaXExSQ}
 \delta X=\frac{1}{2}\varepsilon_{ab}\hat{Q}^a(X)\hat{Q}^b(X)
 \delta F,
\end{eqnarray}
 parameterized by an arbitrary Boson $\delta F$, satisfies
 eq.~(\ref{VarXExSQ}). Furthermore, by analogy with the theorems proved in
 \cite{Sp(2)}, it can be established that any solution of the equations
 (\ref{VarXExSQ}), vanishing when all the variables entering the functional
 $\delta X$ are equal to zero, has the form (\ref{DeltaXExSQ}) with a
 certain Bosonic functional $\delta F$.

 Denote by $Z_X\equiv Z$ the value of the vacuum functional
 (\ref{ZExSQ}) corresponding to the choice of gauge condition
 in the form of the functional $X$. In the vacuum functional
 $Z_{X+\delta X}$ we first make the change of variables
 (\ref{BRSTExSQ}) with $\mu_a=\mu_a(\Phi,\bar\Phi)$, and then
 the additional change of variables
\[
 \delta\Phi^A=(\Phi^A,\delta Y_a)^a,\quad
 \delta\bar{\Phi}_A=(\bar{\Phi}_A,\delta Y_a)^a,\quad\varepsilon(\delta Y_a)=1
\]
 with $\delta Y_a=-i\hbar\mu_a(\Phi,\bar\Phi)$. We get
\[
 Z_{X+\delta X}=\int d\Phi\,d\bar\Phi \rho(\bar\Phi)
 \exp\left\{\frac{i}{\hbar}\bigg(W+X+\delta X+\delta X_1
 +\bar\Phi\Phi\bigg)\right\}\!,
\]
 where
\[
 \delta X_1=2\bigg((X,\delta Y_a)^a-U^a\delta Y_a - i\hbar\Delta^a\delta
 Y_a\bigg)=2\hat{Q}^a(X)\delta Y_a.
\]
 Having eq.~(\ref{DeltaXExSQ}) in mind,
 choose the functional $\delta Y_a$ in the form
\[
 \delta Y_a=-\frac{1}{4}\varepsilon_{ab}\hat{Q}^b(X)\delta F.
\]
 Then we find that $\delta X+\delta X_1=0$ and conclude that the relation
 $Z_{X+\delta X}=Z_X$ holds true. This implies that the symmetry
 transformations (\ref{BRSTExSQ}) do encode the gauge-independence of the
 $S$-matrix within the proposed superfield formalism, and therefore they
 play the role of the transformations of extended BRST symmetry.

\section{Discussion}
\setcounter{equation}{0}

 In this paper we have extended the $Sp(2)$ covariant superfield approach
 \cite{L} to general gauge theories on the basis of fixing the gauge in
 terms of a special generating equation (\ref{GEqXExSQ}) imposed on the
 gauge functional. We have observed that the possibilities of explicit
 representation of the formalism in terms of the operator $\Delta^a$ and
 the extended antibracket $(\,\,,\,)^a$ with the given algebraic properties
 \cite{L} are reduced to two different choices \cite{L,GM}. We have shown
 that only one of  the two possibilities, in fact \cite{L}, is compatible
 with the requirement of extended BRST symmetry (\ref{BRSTExSQ}) realized
 in terms of supertranslations along the Grassmann coordinates of the
 superspace. We have demonstrated that this form of symmetry
 transformations ensures the gauge-independence of the $S$-matrix.

 On the one hand, the quantization rules based on the manifest form of the
 operator $\Delta^a$ and the extended antibracket \cite{L} actually contain
 the superfield $Sp(2)$ covariant scheme \cite{L} as a particular case of
 gauge-fixing. Indeed, any functional
\[
 X(\Phi)=\frac{1}{2}\varepsilon_{ab}U^aU^b F(\Phi),
\]
 parameterized by an arbitrary Boson, $F=F(\Phi)$, is a solution of the
 generating equation (\ref{GEqX1ExSQ}) and represents the exact form of the
 gauge functional used in \cite{L}.

 On the other hand, the proposed method can be considered as
 a superfield form of the modified triplectic approach suggested
 in \cite{mod3pl}. Indeed, consider the component representation
 of superfields $\Phi^A(\theta)$ and supersources $\bar\Phi_A(\theta)$
\begin{eqnarray*}
 \Phi^A(\theta)&=&\phi^A+\pi^{Aa}\theta_a+\frac{1}{2}\lambda^A\theta_a
 \theta^a,\\
 \bar{\Phi}_A(\theta)&=&\bar{\phi}_A-\theta^a\phi^*_{Aa}-\frac{1}{2}\theta_a
 \theta^aJ_A.
\end{eqnarray*}
 The set of variables ($\phi^A, \pi^{Aa}, \lambda^A,\phi^*_{Aa},
 \bar{\phi}_A, J_A$) is identical with the sets of variables
 applied by the $Sp(2)$ covariant \cite{Sp(2)}, triplectic
 \cite{3pl} and modified triplectic \cite{mod3pl} quantization schemes.

 Denote $F(\Phi,\bar{\Phi})\equiv\tilde{F}(\phi,\pi,\lambda,\bar{\phi},\phi^*,J)$.
 Then the component representation of the extended antibracket (\ref{ABExSQ})
\[
 (F,G)^a=\frac{\delta\tilde{F}}{\delta\phi^A}\;\frac{\delta\tilde{G}}
 {\delta\phi^*_{Aa}}+\varepsilon^{ab}\frac{\delta\tilde{F}}{\delta\pi^{Ab}}
 \frac{\delta\tilde{G}}{\delta\bar{\phi}_A}
 -(\tilde{F}\leftrightarrow\tilde{G})\;(-1)^{(\varepsilon(F)+1)(\varepsilon(G)+1)}
\]
 and the operator $\Delta^a$ (\ref{DeltaaExSQ})
\[
 \Delta^a=(-1)^{\varepsilon_A}\frac{\delta_{\it l}}{\delta\phi^A}\;
 \frac{\delta}{\delta\phi^*_{Aa}}+(-1)^{\varepsilon_A+1}\varepsilon^{ab}
 \frac{\delta_{\it l}}{\delta\pi^{Ab}}\;\frac{\delta}{\delta\bar{\phi}_A}
\]
 coincides with the corresponding objects used in \cite{3pl,mod3pl}.
 The form of the integration measure in eq.~(\ref{ZExSQ})
\begin{eqnarray}
\nonumber
 d\Phi\,d\bar{\Phi}\,\rho(\bar{\Phi})=d\phi\,d\phi^*\,d\pi\,d\bar{\phi}\,
 d\lambda\,dJ\,\delta(J)
\end{eqnarray}
 and the component representation of the operator $V^a$ in eq.~(\ref{U&V})
\[
 V^a=\varepsilon^{ab}\phi^*_{Ab}\frac{\delta}{\delta\bar{\phi}_A}-
 J_A\frac{\delta}{\delta\phi^*_{Aa}}
\]
 imply that, when $J_A=0$, the generating equation (\ref{GEq1ExSQ})
 for the action ${\cal W}=W|_{J=0}$ coincides with the one
 used in \cite{mod3pl} when formulating the rules of the modified
 triplectic approach. As for the equation used to define the gauge
 functional $X$ (\ref{GEqX1ExSQ}), note, first of all, that the
 operator $U^a$ (\ref{U&V}), having the component representation
\[
 U^a=(-1)^{\varepsilon_A}\varepsilon^{ab}\lambda^A\frac{\delta_{\it l}}
 {\delta\pi^{Ab}}-(-1)^{\varepsilon_A}\pi^{Aa}
 \frac{\delta_{\it l}}{\delta\phi^A},
\]
 coincides, when $\lambda^A = 0$, with the operator $U^a$
\begin{eqnarray}
\label{UaMTQ}
 U^a=-(-1)^{\varepsilon_A}\pi^{Aa}\frac{\delta_{\it l}}{\delta\phi^A},
\end{eqnarray}
 used in the generating equation that determines the gauge in \cite{mod3pl}.
 Further, note that the functional $\bar\Phi\Phi$ in (\ref{BilFExSQ})
 is given by
\[
 \bar{\Phi}\Phi=\bar{\phi}_A\lambda^A+\phi^*_{Aa}\pi^{Aa}-J_A\phi^A.
\]
 Then we can see that the generating equation for
 the functional ${\cal X} = X + \bar{\phi}_A\lambda^A$
 has the form of eq.~(\ref{GEqX1ExSQ}) with the truncated operator
 $U^a$ (\ref{UaMTQ}), which is formally identical with the generating
 equation of the modified triplectic approach \cite{mod3pl}. As a
 consequence, the vacuum functional
\begin{eqnarray}
\nonumber
 Z=\int d\phi\;d\phi^*\,d\pi\,d\bar{\phi}\,d\lambda\,
 \exp\bigg\{\frac{i}{\hbar}\bigg[{\cal W}(\phi,\pi,\phi^*,\bar{\phi}) +
 {\cal X}(\phi,\pi,\phi^*,\bar{\phi},\lambda) + \phi^*_{Aa}
 \pi^{Aa}\bigg]\bigg\}
\end{eqnarray}
 is identical with the one used in \cite{mod3pl}, limited to the case
 when the action ${\cal W}$ does not depend on the variables $\lambda^A$.

\bigskip

\noindent
{\large
{\bf {Acknowledgments}}}

\vspace{0.5cm}
\noindent
 The authors would like to thank the referee for helpful criticism.
 The work was partially supported by the Russian Foundation  for Basic
 Research (RFBR), project 99-02-16617, as well as by the Russian
 Ministry of Education (Fundamental Sciences Grant E00-3.3-461).
 The work of  P.M.L. was also supported by INTAS, grant 99-0590,
 and by  the joint project of RFBR and Deutsche Forschungsgemeinschaft
 (DFG), 99-02-04022.

\vspace{0.5 cm}

\noindent
{\Large{\bf{Appendix A}}}
\setcounter{equation}{0}
\renewcommand{\theequation}{A.\arabic{equation}}

\bigskip

\noindent
 Consider a set of linearly independent Fermionic doublets
 $D^a=(\Delta^a,U^a,V^a)$ subject to eqs.~(\ref{algUV}) and
 (\ref{algDelta}). Let us introduce two Bosonic objects $C^{(1,2)}$ by the
 rule
\begin{eqnarray*}
 C^1&=&\{U^1,\Delta^2\}=-\{U^2,\Delta^1\}\not\equiv 0,\\
 C^2&=&\{V^1,\Delta^2\}=-\{V^2,\Delta^1\}\not\equiv 0.
\end{eqnarray*}
 It is straightforward to check that $C^{(1,2)}$ commute
 with $D^a$, and, consequently, also with each other
\begin{eqnarray}
\label{comm}
 [C^{(1,2)},D^a]=[C^1,C^2]=0.
\end{eqnarray}
 In general, the objects $C^{(1,2)}$ may be linearly dependent, i.e.
\begin{eqnarray}
\label{i}
 &&C^1=C^2=0,\\
\label{ii}
 &&C^1=\alpha C^2\neq 0,\\
\label{iii}
 &&C^1=0,\,\,C^2\neq 0,\\
 \label{iv}
 &&C^1\neq 0,\,\,C^2=0
\end{eqnarray}
 and linearly independent, i.e.
\begin{eqnarray}
\label{v}
 &&C^1\neq 0,\,\,C^2\neq 0,\,\,
 C^2\neq\alpha C^1.
\end{eqnarray}
 The above possibilities imply, by virtue of eqs.~(\ref{algUV}),
 (\ref{algDelta}), (\ref{comm}), that the whole set of Fermions $D^a$ and
 Bosons $C^{(1,2)}$ generally spans five different Lie superalgebras: a
 six-dimensional one (\ref{i}), three seven-dimensional ones
 (\ref{ii})--(\ref{iv}), and an eight-dimensional one (\ref{v}).

 From eqs.~(\ref{algUV}), (\ref{algDelta}) and (\ref{comm})
 it follows that each of the resulting superalgebras ${\cal G}$
 is nilpotent (see, e.g., \cite{dict}) with respect to the
 supercommutator $[\,\,,\,\}$, namely, for the sequence
 ${\cal G}^{[i]}$
\[
 [{\cal G},{\cal G}\}={\cal G}^{[1]},\,\,\,
 [{\cal G},{\cal G}^{[i-1]}\}={\cal G}^{[i]},\,\,\,
 {\cal G}\equiv{\cal G}^{[0]},\,\,\,
 i\geq 1
\]
 there exists an integer $n$ such that ${\cal G}^{[n]}=\{0\}$. Thus, in the case
 (\ref{i}), we have $n=1$, while in the other cases (\ref{i})--(\ref{v}),
 $n=2$. On the other hand, all the given superalgebras possess a non-trivial
 ideal \cite{dict}, namely, there exists a non-trivial subalgebra ${\cal I}$
\[
 {\cal I}\subset{\cal G},\,\,\,
 {\cal I}\neq\{0\},\,\,
 {\cal I}\neq{\cal G},\,\,\,
 [{\cal I},{\cal I}\}\subset{\cal I}
\]
 such that $[{\cal G},{\cal I}\}\subset{\cal I}$.
 Thus, for the superalgebra (\ref{i})
 the ideal ${\cal I}$ is given by the linear span of any five
 elements from the set ($U^a$, $V^a$, $\Delta^a$), and for the
 remaining cases (\ref{ii})--(\ref{v}), by the linear span of
 $C^{(1,2)}$. The existence of a non-trivial ideal ${\cal I}$
 implies that the above superalgebras do not belong to simple
 Lie superalgebras, described by the standard classification
 \cite{dict}. Note also that these superalgebras are not
 semi-simple \cite{dict} either, because the ideal ${\cal I}$
 is solvable, i.e. for the sequence ${\cal I}^{(i)}$
\[
 [{\cal I},{\cal I}\}={\cal I}^{(1)},\,\,\,
 [{\cal I}^{(i-1)},{\cal I}^{(i-1)]}\}={\cal I}^{(i)},\,\,\,
 {\cal I}\equiv{\cal I}^{(0)},\,\,\,
 i\geq 1
\]
 there exists an integer $n$ such that ${\cal I}^{(n)}=\{0\}$. Thus, in all
 cases (\ref{i})--(\ref{v}), we have $n=1$.

 Let us turn to the manifest representation of the operators $U^a$ and
 $V^a$ as generators of supertranslations (\ref{U&V}) and consider the
 explicit choices $\Delta^a_1$, $\Delta^a_2$ for the operator $\Delta^a$ in
 the form (\ref{DeltaaExSQ}), (\ref{DeltaaExSQ2}), respectively. It is
 straightforward to check that the cases (\ref{DeltaaExSQ}) and
 (\ref{DeltaaExSQ2}) lead to different superalgebras. Thus, in the case
 (\ref{DeltaaExSQ}) we have a realization of the form (\ref{ii}), with
\begin{eqnarray}
\label{manif}
 C^1=-C^2=\int d^2\theta\,\frac{\delta_l}{\delta\Phi(\theta)}
 \frac{\partial^2}{\partial\theta^2}\frac{\delta}{\delta\bar{\Phi}(\theta)}
 (-1)^{\varepsilon_A},
\end{eqnarray}
 while in the case (\ref{DeltaaExSQ2}) we arrive at the realization
 (\ref{iii}), with $C^2$ given by (\ref{manif}). Finally, note that
 non-vanishing linear combinations of the operators $\Delta^a_1$,
 $\Delta^a_2$ are restricted to the seven-dimensional superalgebras
 (\ref{ii})--(\ref{iv}).

\end{document}